\documentclass[reprint,superscriptaddress,groupedaddress,aps,prb,floatfix,longbibliography]{revtex4-2}
\usepackage{xcolor}
\usepackage[hypertexnames=false]{hyperref}
\hypersetup{
    colorlinks=true, citecolor={blue},
    linkcolor={blue}, urlcolor={blue},filecolor={blue}}
\usepackage{graphicx}
\usepackage{amsmath}
\usepackage{textcomp, gensymb}
\usepackage{amssymb}

\usepackage[T1]{fontenc}
\usepackage{lmodern}            
\usepackage[final]{microtype}

\usepackage{siunitx}
\usepackage{dcolumn}
\usepackage{float}
\usepackage{bm}

\begin{document}


\title{Hanle Lineshapes and Spin-Rotation Signatures from In-Plane Anisotropic Spin Relaxation in Heterogeneous Spin Devices}

\author{Josef Sv\v{e}tlík$^{1,2}$}
\email{josef.svetlik@icn2.cat}
\author{Juan F. Sierra$^{1}$}
\author{Lorenzo Camosi$^{1}$}
\author{Williams Savero Torres$^{1}$}
\author{Franz Herling$^{1}$}
\author {Vera Marinova$^{3}$}
\author{Dimitre Dimitrov$^{3,4}$}
\author{Sergio O. Valenzuela$^{1,5}$}
 \email{SOV@icrea.cat}

\affiliation{
$^{1}$Catalan Institute of Nanoscience and Nanotechnology (ICN2), CSIC and The Barcelona Institute of Science and Technology (BIST), Campus UAB, Bellaterra, 08193 Barcelona, Spain
}%
\affiliation{$^2$Universitat Aut\`{o}noma de Barcelona, Department of Physics, Bellaterra, 08193 Barcelona, Spain}
\affiliation{$^3$Institute of Optical Materials and Technologies, Bulgarian Academy of Science, 1113 Sofia, Bulgaria}
\affiliation{$^4$Institute of Solid State Physics, Bulgarian Academy of Sciences, 1784 Sofia,  Bulgaria}
\affiliation{$^5$Instituci\'{o} Catalana de Recerca i Estudis Avan\c{c}ats (ICREA), 08010 Barcelona, Spain}

\date{\today}

\begin{abstract}

Spin precession experiments in lateral spin devices are a powerful tool for probing the spin transport properties of materials. These experiments can be quantitatively described using the Bloch diffusion equation, which offers a practical framework for modeling spin-related phenomena. In this work, we present calculations of the spin density across heterogeneous, diffusive spintronic devices. The modeled devices feature spin transport channels that include both isotropic and in-plane anisotropic spin relaxation regions. We analyze how different geometric configurations and spin transport parameters influence the lineshape of spin precession signals under magnetic fields applied in different orientations and compare with experimental observations. Our results introduce a theoretical framework for interpreting spin-transport measurements in lateral graphene spin devices. The framework is especially relevant when the graphene is partially proximitized by other two-dimensional materials, where proximity-induced spin–orbit coupling leads to anisotropic spin relaxation.

\end{abstract}

\maketitle


\section{\label{sec:Introduction}Introduction}
Two-dimensional materials (2DMs) have emerged as promising platforms for spintronics due to their unique electronic properties, atomically thin structure, and pronounced spin-related phenomena \citep{valenzuela_spintronic_2024}. Materials such as graphene, transition metal dichalcogenides (TMDCs) and their combination in complex heterolayers, offer long spin diffusion lengths \cite{han_graphene_2014,drogeler_2016,Gebeyehu_2019}, high carrier mobility, and the potential for gate-tunable spin–orbit coupling (SOC) \citep{zutic_proximitized_2019,sierra_van_2021}. In particular, hybrid systems comprising single-layer graphene (SLG) and 2DMs with strong SOC—especially TMDCs—have attracted increasing interest due to the wide range of available material combinations and the possibility to induce SOC in graphene of several meV \citep{wang_strong_2015,wang_origin_2016,wakamura_strong_2018}. Moreover, the resulting spin–orbit fields (SOFs) emerging in graphene can be engineered in various spatial directions \citep{naimer_twist-angle_2021,zollner_first-principles_2025}. In stark contrast to the well-established observation of isotropic spin transport in conventional SLG spin devices on SiO$_2$ \citep{raes_determination_2016,raes_spin_2017}, hybrid SLG-TMDC systems exhibit pronounced anisotropic spin transport. The most widely studied example involves SLG combined with TMDCs in the 2H phase, such as WS$_2$, MoS$_2$, MoSe$_2$, and WSe$_2$ \citep{ghiasi_large_2017,benitez_strongly_2018,herling_gate_2020,hoque_spin-valley_2023}. These heterostructures, which have an inherent three-fold symmetry, present anisotropic spin transport driven by a dominant out-of-plane SOF, arising from the valley-Zeeman coupling \citep{gmitra_graphene_2015,gmitra_trivial_2016}. As a result, the lifetime of spins oriented perpendicular to the plane of the heterostructure, denoted $\tau_{z}$, exceeds that of the in-plane spin components, $\tau_{x}$ and $\tau_{y}$, which are equal due to symmetry considerations \citep{cummings_giant_2017,garcia_spin_2018}. More recently, a strong spin lifetime anisotropy between two orthogonal in-plane directions has been experimentally observed in heterostructures comprising graphene and pentagonal PdSe$_2$ \citep{sierra_room-temperature_2025}. In this system, the intrinsic anisotropic properties in the plane of PdSe$_2$ give rise to a dominant SOF in the plane, leading to a pronounced variation in the spin lifetimes oriented along $x$ and $y$, \textit{i.e.}, $ \tau_x \not= \tau_y$. Similar results have been found theoretically with low-symmetry SnTe \cite{Milivojevic_giant_2024}. These findings demonstrate that the symmetry of the heterostructure plays a key role in shaping the effective SOF orientation and, consequently, the spin relaxation anisotropy.

\begin{figure} [ht]
\centering
\includegraphics[width=0.8\linewidth]{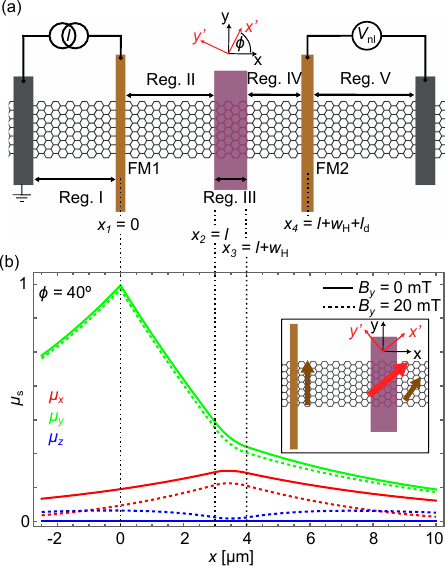}
\caption{(a) Schematics of a heterogeneous lateral spin device comprising one anisotropic and four isotropic regions (Reg. I-V). Region III denotes the anisotropic region, where spins oriented along the three spatial directions can exhibit different lifetimes. A pure spin current flows between the injector (FM1) and detector (FM2) ferromagnetic electrodes. The injected spins, that then diffuse along $x$, are initially aligned with $y$, the easy magnetization axis of FM1 and FM2. The red arrows, labeled $x'$ and $y'$, denote the principal axes of spin relaxation, which do not need to coincide with $x$ and $y$; they are rotated by an angle $\phi$ with respect to the device axes. An external magnetic field $B$ is used to induce spin precession and vary the orientation of the spins in region III. (b) Spin-dependent electrochemical potentials $\mu_x$, $\mu_y$, $\mu_z$ across the device for $\phi=40^\circ$, with $B_y=0$ (solid curves) and $B_y=20 \ $mT (dashed curves). Although spins are injected along $y$, anisotropic relaxation yields a nonzero $\mu_x$ even at $B_y=0$. Applying $B_y$ induces spin precession, as spins are no longer parallel to $y$, which leads to the generation of an additional $\mu_z$ component (dashed blue curve). Inset: Schematic representation of the direction of dominant SOF  (large red arrow) and the spin orientation at the injection point and after diffusing through the anisotropic region (brown arrows).}
 \label{fig:RegMod}
\end{figure}

A well-established technique for studying spin transport involves investigating spin precession under applied magnetic fields. In the standard (Hanle) configuration, the magnetic field is applied perpendicular to the plane of the device, causing spins to precess within the plane \citep{johnson_interfacial_1985,jedema_electrical_2002}. However, this configuration does not allow for distinguishing spin lifetimes along all spatial directions. To obtain this information, it is necessary to apply the magnetic field in a configuration that drives the spins out-of-plane \citep{raes_determination_2016,raes_spin_2017,li_oblique_2008,motsnyi_optical_2003}. This can be achieved, for instance, by tilting the magnetic field away from perpendicular and towards the magnetization of the ferromagnetic injector (the oblique configuration \citep{raes_determination_2016}) or by applying the magnetic field in-plane, perpendicular to the magnetization \citep{raes_spin_2017}.  Such approaches have been implemented in nonlocal spin devices based on SLG \citep{raes_determination_2016,ghiasi_large_2017,benitez_strongly_2018, hoque_spin-valley_2023}, allowing the determination of spin lifetimes for spins oriented along different spatial directions, which plays a crucial role in determining the orientation of the dominant SOFs \citep{raes_spin_2017}. Figure$ \ $\ref{fig:RegMod} shows the heterogeneous device geometry considered in this article, along with a schematic of the nonlocal measurement configuration, where the voltage probes are spatially separated from the current path. This arrangement enables the generation of a pure spin current that diffuses from the spin ferromagnetic injector (FM1) to the detector (FM2) and significantly suppresses charge-related contributions to the detected signal \citep{johnson_interfacial_1985,jedema_electrical_2002,tombros_electronic_2007,valenzuela_nonlocal_2009,han_graphene_2014}. In the oblique magnetic field configuration, the external magnetic field $B$ is applied within a plane defined by the direction perpendicular to the substrate and the easy axis of the ferromagnetic spin injector and detector, which is set by the shape anisotropy of the contacts, \textit{i.e.}, aligned with their long axis along $y$ \citep{raes_determination_2016,raes_spin_2017}. The orientation of the oblique field is defined by the angle $\beta$, measured from the substrate toward the out-of-plane direction.
Spin lifetimes $\tau_{x},\tau_{y},\tau_{z}$ can be extracted by fitting the oblique spin precession curves with the analytical solution of the Bloch diffusion equation \citep{torrey_bloch_1956}, solved for the system defined by a specific device geometry and boundary conditions, allowing the determination of the spin anisotropy parameter $\zeta_{ij}=\tau_i/\tau_j$ , defined as the ratio between the spin lifetimes for spins oriented along $i$ and $j$ \citep{raes_determination_2016,raes_spin_2017}.

In this article, we construct solutions to the Bloch diffusion equation for the device in Fig.$ \ $\ref{fig:RegMod}, which comprises isotropic regions intercalated with an anisotropic region that models a hybrid SLG–TMDC heterobilayer exhibiting strong in-plane spin anisotropy. This scenario is fundamentally different from the out-of-plane Zeeman-type spin-orbit fields characteristic of 2H-TMDCs, which instead lead to isotropic in-plane spin relaxation \citep{benitez_investigating_2019}. In the isotropic regions, spin relaxation is described by the spin lifetime $\tau_\mathrm{s}$, independent of spin orientation, as is the case for the pristine SLG on SiO$_2$ substrates. \citep{raes_determination_2016,raes_spin_2017}. In contrast, the anisotropic region exhibits orientation-dependent spin relaxation, where the spin population decays at different rates along three orthogonal directions. We show that, in such heterogeneous channels, the spin diffusion through isotropic segments coupled to an anisotropic region produces characteristic precession signatures.

The article is organized as follows. In Section \ref{sec:Bloch_eq}, we present the general solution to the Bloch diffusion equation in heterogeneous spin devices. We then analyze the resulting spin density profile across the device (Section \ref{sec:spin_density}). Our results show that, in the presence of strong in-plane spin anisotropy, an effective rotation of the injected spins as they traverse the anisotropic region can lead to a distinctive decay in the spin signal when $B$ is applied along the easy axis of the spin injector/detector, when no torque is exerted on the spins in the isotropic regions (Section \ref{sec:inplane}). This behavior, which has not been described before, serves as a robust signature of in-plane spin anisotropy and can be used as a diagnostic tool in experimental studies. In Section \ref{sec:l_ld_dep}, we also show that when $B$ is applied out of plane—so that spins precess exclusively within the plane—the resulting spin precession signal can exhibit a pronounced asymmetry in the lobes corresponding to the collective $\pi$ precession of the spins at the detection point. This asymmetry is another strong indication of in-plane anisotropy. The behavior of the spin signal is further investigated under different oblique magnetic field orientations, illustrating the effect of out-of-plane spin lifetime on the precession lineshape (Section \ref{sec:simul-obl}). In the final section (Section \ref{sec:exp}), we illustrate how the modeled results compare with the experimental data obtained in a graphene-PdSe$_2$ device.

\section{\label{sec:Bloch_eq}Solution of the Bloch diffusion equation}

The behavior of spins as they diffuse, precess, and relax is captured by the diffusive Bloch equation,
\begin{equation} \label{Eq:Bloch}
D_\mathrm{s} \, \Delta \, \vec{\mu'}_\mathrm{s} -\gamma _c \ \vec{\mu'}_\mathrm{s} \times \vec{B'} - \overline{(\tau '_\mathrm{s})^{-1}} \cdot \vec{\mu'}_\mathrm{s} = 0 ,
\end{equation}
where $D_\mathrm{s}$ is the spin diffusion constant, $\gamma_c$ is the gyromagnetic ratio, $\vec{\mu'}_\mathrm{s}=(\mu_{x'} , \mu_{y'}, \mu_z)$ is the spin-dependent electrochemical potential, $\vec{B'}$ is the magnetic field, and $\overline{(\tau '_{s})^{-1}}$ is the diagonal matrix containing the spin relaxation times along the three orthogonal directions,
\begin{equation}
\overline{(\tau '_\mathrm{s})^{-1}} = \begin{pmatrix} (\tau _{x'})^{-1} & 0 & 0 \\0 & (\tau _{y'})^{-1} & 0 \\ 0 & 0 & (\tau _{z})^{-1} \end{pmatrix}
\end{equation}
The equation is presented in the coordinate system defined by $\hat{x}'$ and  $\hat{y}'$, which is rotated by an angle $\phi$ within the plane relative to the original coordinate axes $\hat{x}$ and $\hat{y}$, aligned with the spin propagation direction and the direction of the injected spins, respectively. The relation between the two coordinate systems is given by:
\begin{equation}
    \begin{aligned}
        x'&= x \ \mathrm{cos}(\phi) + y \ \mathrm{sin}(\phi) \\
        y'&= -x \ \mathrm{sin}(\phi) + y \ \mathrm{cos}(\phi) ,
    \end{aligned}
\end{equation}
This rotated coordinate system introduces the possibility of a misalignment between the direction of the injected spins $\hat{y}$ and the principal axes of spin relaxation within the heterogeneous region—namely, the directions along which spins experience the longest (defined as $\hat{x}'$) and the shortest (defined as $\hat{y}'$) lifetimes \citep{sierra_room-temperature_2025}. In practice, in low-symmetry systems, the angle $\phi$ that defines this rotation is typically unknown prior to the experiments, especially if the crystalline orientations of the SLG and TMDC remain undetermined. In the new coordinate system, the injected spin current $\vec{J}_{s_0}=(0 ,  J_{y_0} ,  0)$ is expressed as:
\begin{equation} \label{Eq:rotInjection}
      \vec{J_{\mathrm{s}_0}'} =\Bigl(J_{x_0'}  , J_{y_0'}  , 0 \Bigl)
      =\Bigl(J_{y_0} \ \mathrm{sin}(\phi),  J_{y_0} \ \mathrm{cos}(\phi) , 0\Bigl)
\end{equation}
with the oblique magnetic field applied at an angle $\beta$ within the $yz$ plane, $\vec{B}=\bigl(0\ , \,  B \ \mathrm{cos}(\beta)\ , \,  B \ \mathrm{sin}(\beta)\bigl)$ as:
\begin{equation}
     \vec{B'} = 
     \Bigl(B \ \mathrm{sin}(\phi) \ \mathrm{cos}(\beta)  , B \ \mathrm{cos}(\phi) \ \mathrm{cos}(\beta) ,  B \ \mathrm{sin}(\beta)\Bigl) .
\end{equation}
In spin devices consisting of a long, narrow channel oriented along $\hat{x}$, the spin transport is approximately one-dimensional, and the equation \ref{Eq:Bloch} can be rewritten as:
\begin{equation} \label{Eq:compact}
   D_\mathrm{s} \, \frac{\partial^2}{\partial x^2} \, \vec{\mu'}_\mathrm{s} =
    \begin{pmatrix}
        (\tau _{x'})^{-1} & - \gamma_c \, B_z &  \gamma_c \, B_y'\\  \gamma_c \, B_z& (\tau _{y'})^{-1} & - \gamma_c \, B_x' \\ - \gamma_c \, B_y' &  \gamma_c \, B_x' & (\tau _{z})^{-1}
    \end{pmatrix} \, \vec{\mu'}_\mathrm{s} .
\end{equation}
The mathematical solution of this equation has the form $\vec{\mu'}_\mathrm{s}= e^{k \, x} \ \vec{v}$, where $k$ is a constant and $\vec{v}$ a vector. The second derivative of Eq.$ \ $\ref{Eq:compact} after substitution is:
\begin{equation} 
    D_\mathrm{s} \ k^2 \ e^{k \, x} \ \vec{v}= A \ e^{k \, x} \ \vec{v} ,
\end{equation}
where A is the matrix in Eq.$ \ $\ref{Eq:compact}. It can be reduced to
\begin{equation} \label{Eq:eigen}
    (A-\lambda \ I) \ \vec{v}=0 ,
    \end{equation}
where $\lambda=D_\mathrm{s} \ k^2$ and $I$ is the $3\times3$ identity matrix. For every root of Eq.$ \ $\ref{Eq:eigen}, there is a linearly independent solution of the form $\vec{\mu'}_\mathrm{s}= e^{k \, x} \ \vec{v}$. A linear combination of all independent solutions provides a general solution,
\begin{equation}
    \vec{\mu'}_\mathrm{s}= \sum_{n=1}^{3} \Bigl(c_n^+ \ e^{k_n^+ \ x}+c_n^- \ e^{k_n^- \ x}\Bigl) \ \vec{v}_n ,
    \end{equation}
where $k_n^\pm=\pm \sqrt{\lambda_n/D_\mathrm{s}}$.  $\lambda_n$ and $\vec{v}_n$ are eigenvalues and eigenvectors of $A$.
The constants $c_n^\pm$ can be determined by setting the boundary conditions.

The boundary conditions for the heterogeneous spin valve, characterized by the five different regions [see Fig.$ \ $\ref{fig:RegMod}(a)], are:
\begin{equation}
 \begin{aligned}
      \vec{\mu'}^m_\mathrm{s}(x_m) &=\vec{\mu'}^{m+1}_\mathrm{s}(x_m)
 \\ \\   \frac{\mathrm{d} \vec{\mu'}^m_\mathrm{s}}{\mathrm{d}x}\Bigg|_{x=x_m} &= \frac{\mathrm{d} \vec{\mu'}^{m+1}_\mathrm{s}}{\mathrm{d}x}\Bigg|_{x=x_m} +
 \begin{cases}
 \vec{J_{\mathrm{s}_0}'}, & m=1 \\
 \ 0,& m \neq 1 ,

 \end{cases}
 \end{aligned}
\end{equation}
where $x_m$ is the intersection of regions $m$ and $m+1$. Regions I, II, IV, and V correspond to pristine graphene, characterized by isotropic spin transport ($\tau_{x'} = \tau_{y'} = \tau_z$) \citep{raes_determination_2016}. Region III is graphene proximitized with an insulating 2DM of width $w_\mathrm{H}$, which induces anisotropic spin transport in graphene ($\tau_{x'} \neq \tau_{y'} \neq \tau_z$). For simplicity, we assume that $D_\mathrm{s}$ is invariant across the device. FM1 generates a spin current $\vec{J}_{s_0}$ at the interface of regions I and II ($x_1=0$), which diffuses into both sides. The injected spin current has a magnitude $J_{y_0} \propto {P_\mathrm{i} \ J_\mathrm{c}}$, where $P_\mathrm{i}$ is FM1's spin polarization and $J_\mathrm{c}$ the applied charge current.

\begin{figure*} [ht]
\centering
\includegraphics[width=0.9\linewidth]{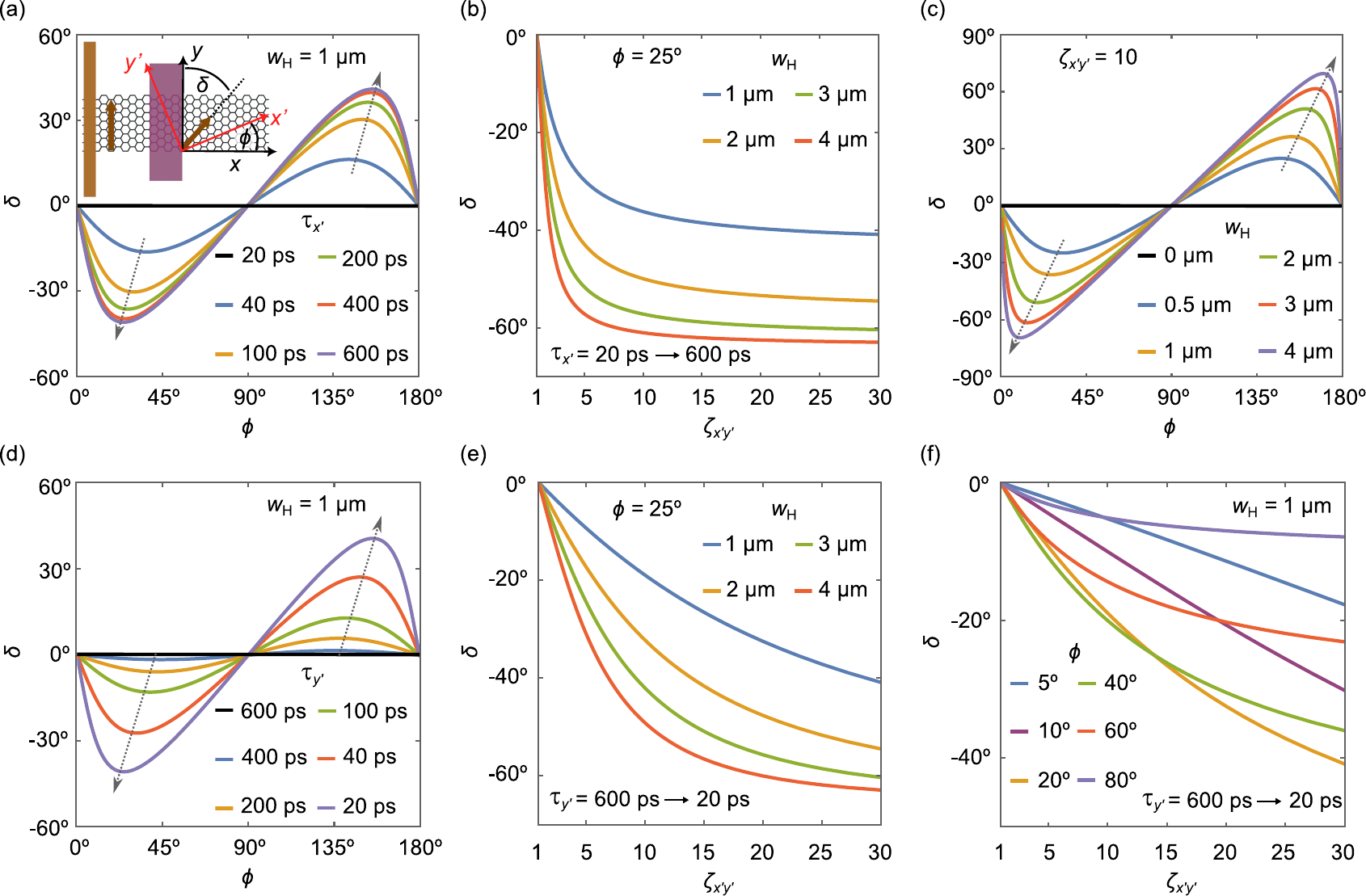}
\caption{
Spin rotation in the anisotropic region. (a) $\delta$ vs. $\phi$ for different $\tau_{x'}$. Inset: definition of $\delta$. In the limit where the $y'$ spin component is fully suppressed ($\tau_{y'}<\tau_{x'}$ by definition), spins align with $x'$, yielding $\delta = 90\degree-\phi$. (b) $\delta$ vs. $\zeta_{x'y'}$ for different $w_\mathrm{H}$ at $\phi=25\degree$. In (a) and (b) $\zeta_{x'y'}$ is modulated by varying $\tau_{x'}$ and keeping $\tau_{y'}$ fixed at 20 ps. (c) $\delta$ vs. $\phi$ for different $w_\mathrm{H}$. As $w_\mathrm{H}$ increases, the $y'$ spin component decays exponentially faster than the $x'$ component, and $\delta$ approaches $90\degree-\phi$.  (d) $\delta$ vs. $\phi$ for different $\tau_{y'}$. (e)$\, \delta \,$vs. $\zeta_{x'y'}$ for different $w_\mathrm{H}$. In (d) and (e) $\zeta_{x'y'}$ is modulated by varying $\tau_{y'}$ and keeping $\tau_{x'}$ fixed at 600 ps. (f) $\delta$ vs. $\zeta_{x'y'}$ for different $\phi$. The values of all the other parameters are listed in Table$\ $\ref{Box:Par1}.}
 \label{fig:SOF_gate}
\end{figure*}

After calculating $\vec{\mu'}_\mathrm{s}$ across the sample, an inverse transformation
\begin{equation}
    \begin{aligned}
        x&= x' \ \mathrm{cos}(\phi) - y \ \mathrm{sin}(\phi) \\
        y&= x' \ \mathrm{sin}(\phi) + y \ \mathrm{cos}(\phi) \
    \end{aligned}
\end{equation}
is performed to obtain $\vec{\mu}_\mathrm{s}$ in the original coordinate system
\begin{equation}
    \vec{\mu}_\mathrm{s}
    =\Bigl({\mu}_{x'} \ \mathrm{cos}(\phi)-{\mu}_{y'} \ \mathrm{sin}(\phi) , {\mu}_{x'} \ \mathrm{sin}(\phi)+{\mu}_{y'} \ \mathrm{cos}(\phi) ,  \mu_z \Bigl) .
\end{equation}
The detector FM2, located at the interface between regions IV and V at $x_4=l+w_\mathrm{H}+l_\mathrm{d}=L$, is sensitive only to the $\mu_y$ component, which is collinear with FM2 magnetization. Therefore,
\begin{equation} \label{Eq:detector}
    V_\mathrm{nl}
    =\frac{P_\mathrm{d} \ \Bigl({\mu}_{x'}(L) \ \mathrm{sin}(\phi)+{\mu}_{y'}(L) \ \mathrm{cos}(\phi)\Bigl)}{e} .
\end{equation}
Experimentally, $V_\mathrm{nl}$ is usually normalized by the injected current $I$, yielding the nonlocal resistance defined as $R_\mathrm{nl}=V_\mathrm{nl}/I$.

\section{\label{sec:spin_density}Spin density across heterogeneous spin devices}
Figure$ \, $\ref{fig:RegMod}(b) shows the calculated $\mu_x, \ \mu_y$, and $\mu_z$ across the heterogeneous spin device when $\phi= 40 \degree$. The device dimensions and the considered spin parameters in both the isotropic and anisotropic regions are listed in the Table$\ $\ref{Box:Par1}. The values correspond to our experimental geometries and spin transport parameters, as determined for pristine graphene and for graphene proximitized by a TMDC \cite{raes_determination_2016,benitez_strongly_2018,sierra_room-temperature_2025,benitez_tunable_2020}.

\begin{table}[b]
\setlength{\tabcolsep}{5.9pt}
\renewcommand{\arraystretch}{1.2}
    \centering
    \begin{tabular}{ c c c c }
    \hline
     \hline
          $ l= 3 \ \micro \mathrm{m}$&$ w_\mathrm{H}= 1 \ \micro \mathrm{m}$&   $ L= 11 \ \micro \mathrm{m}$&   $D_\mathrm{s}= 0.05 \ \mathrm{m^2 s^{-1}}$\\

          $\tau_\mathrm{s} = 1 \ \mathrm{ns}$&$ \tau_{x'} = 200 \ \mathrm{ps}$&  $\tau_{y'}= 20 \ \mathrm{ps}$& $\tau_{z}= 5 \ \mathrm{ps}$ \\
         \hline
          \hline
         \end{tabular}
    \caption{Geometric and spin transport parameters used to calculate spin density across the device shown in Fig.$ \ $\ref{fig:RegMod}.}
    \label{Box:Par1}
\end{table}

Interestingly, in the anisotropic region, spins undergo an effective rotation toward $\hat{x}'$ [see the inset of Fig.$ \ $\ref{fig:RegMod}(b)] due to the relatively rapid relaxation of the $\hat{y}'$-oriented component compared to the relaxation of the component along $\hat{x}'$. This anisotropic relaxation leads to a non-zero $\mu_x$, which would not be present in an isotropic system ($\zeta_{x'y'}= 1$) as spins are injected along $y$. Notably, the magnitude of $\mu_x$ becomes comparable to that of $\mu_y$. The generated spin component along $x$ diffuses away from region III, both towards FM1 and FM2. Consequently, applying an external magnetic field $B_y$, which is no longer parallel to the propagating spins, induces an out-of-plane precession that generates a nonzero $\mu_z$, which increases as spins diffuse away from region III [dashed blue line in Fig.$ \ $\ref{fig:RegMod}(b)].

The degree to which spins exiting the anisotropic region are effectively rotated depends on the angle $\phi$, the spin lifetimes $\tau_{x'}, \tau_{y'}$, and the width $w_\mathrm{H}$ of the anisotropic region. Naturally, spins maintain their original orientation even in the presence of anisotropy when $\phi =\pi/2 \ n$, where $n$ is an integer. Spin rotation occurs only when conditions $\zeta_{x'y'} > 1$ and $\phi \neq \pi/2 \ n$ are fulfilled simultaneously. The magnitude of spin rotation increases with both $\zeta_{x'y'}$ and $w_\mathrm{H}$, as the spin component aligned with $\hat{y}'$ undergoes a faster relaxation over a wider region of the device. Figure$ \, $\ref{fig:SOF_gate} shows the effective spin rotation, quantified by the angle $\delta$, which is measured from the orientation of the injected spins [see the inset of Fig.$ \ $\ref{fig:SOF_gate}(a)], as a function of these parameters. In Fig.$ \ $\ref{fig:SOF_gate}(a) $\delta$ vs. $\phi$ is plotted for selected $\tau_{x'}$ when $\tau_{y'}=20 \,$ps and $w_\mathrm{H}=1 \, \micro$m. The angle $\phi_\mathrm{max}$ at which the maximum rotation occurs depends on $\tau_{x'}$ and $\tau_{y'}$ and here shifts from about $40\degree$ to $20\degree$ as $\zeta_{x'y'}$ increases.  Figure$ \, $\ref{fig:SOF_gate}(b) shows $\delta$ as a function of $\zeta_{x'y'}$ for various $w_\mathrm{H}$ values with $\phi=25 \degree$. As in Fig.$ \ $\ref{fig:SOF_gate}(a), the anisotropy is tuned by varying $\tau_{x'}$ while keeping $\tau_{y'}$ fixed at 20 ps. In this situation, $\delta$ changes significantly in the range $\zeta_{x'y'}$ $\in$ [1,10], and tends to saturate for larger anisotropy ratios. For sufficiently large $w_\mathrm{H}$ and $\zeta_{x'y'}$, the component along $y'$ is largely suppressed, and thus $\delta$ is such that spins are along $x'$, that is, $\delta \sim \phi-90$.
Figure$ \, $\ref{fig:SOF_gate}(c) shows $\delta$ vs. $\phi$ for different values of $w_\mathrm{H}$, with fixed anisotropy $\zeta_{x'y'}=10 \, (\tau_{x'}=200 \, \mathrm{ps}, \tau_{y'}= 20 \, \mathrm{ps})$, and demonstrates how $\phi_\mathrm{max}$ shifts toward lower values as $w_\mathrm{H}$ increases. It is observed that, in the limit of $w_\mathrm{H} \rightarrow \infty$,  $\delta \rightarrow  \phi-90$. Figures$ \, $\ref{fig:SOF_gate}(d) and (e) show $\delta$ vs. $\phi$ and $\delta$ vs. $\zeta_{x'y'}$, respectively—similar to Figs. \ref{fig:SOF_gate}(a) and (b)—but in this case, $\tau_{y'}$ is varied while keeping $\tau_{x'}$ fixed at 600 ps. Varying $\zeta_{x'y'}$ by decreasing $\tau_{y'}$ leads to a more gradual change in $\delta$, and saturation occurs for larger $\zeta_{x'y'}$. Finally,  Fig.$ \ $\ref{fig:SOF_gate}(f) shows that a nearly linear modulation of $\delta$ can be achieved in a large range of $\zeta_{x'y'}$ by tuning $\tau_{y'}$ when $\phi$ is small ($\phi \leq 10^\degree$).

These calculations demonstrate that the direction of spin polarization can be significantly rotated away from the spin injection axis as a result of in-plane anisotropic spin relaxation. Furthermore, spin orientation can be controlled by tuning $\zeta_{x'y'}$, which, in certain material systems, can be modulated by electrostatic gating \citep{sierra_room-temperature_2025}. Tuning the SOF (and thus $\zeta_{x'y'}$) via an external electric field offers a promising route for dynamically controlling spin orientation. However, such spin manipulation through anisotropic spin relaxation presents a considerable limitation: it leads to a reduction in the overall magnitude of the spin signal, which is substantial for larger $\delta$.

As described in the following sections, the effective spin rotation within the anisotropic region gives rise to distinctive features in the $R_\mathrm{nl}$ vs. $B$ curves, enabling a clear identification of the in-plane spin relaxation anisotropy with spin precession experiments.

\begin{figure} [!ht]
\centering
\includegraphics[width=0.99\linewidth]{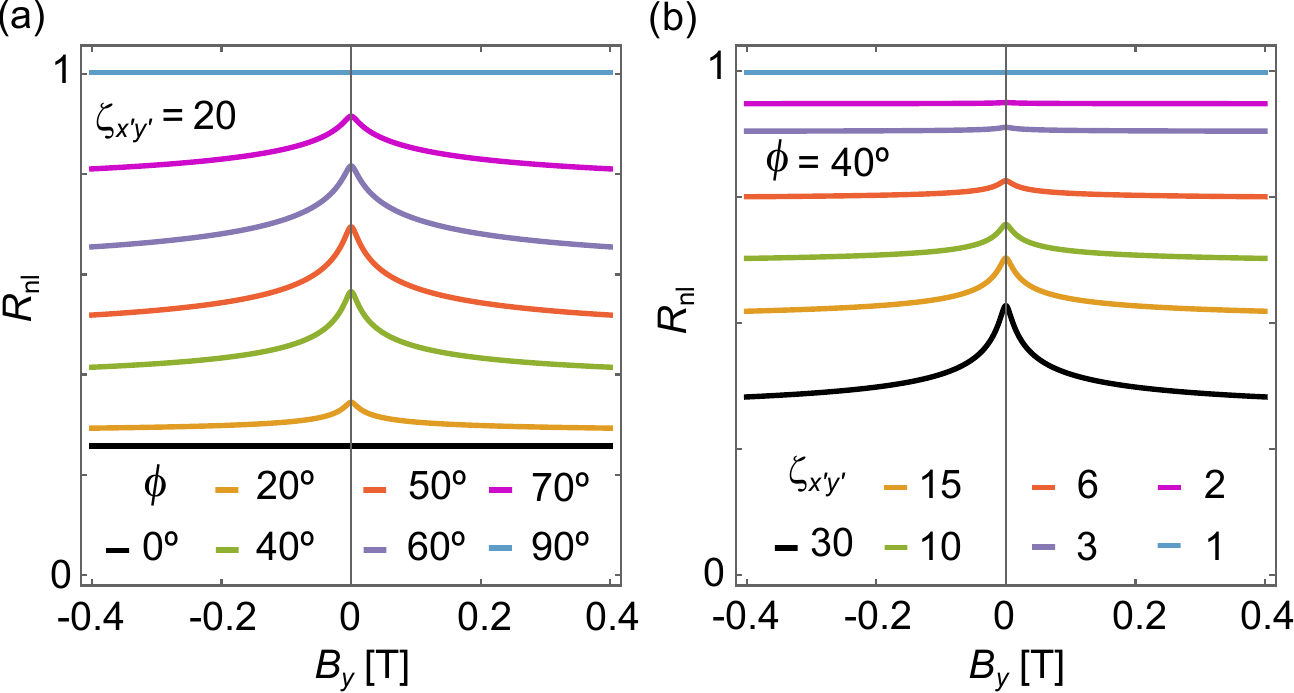}
\caption{Dependence of the nonlocal resistance $R_\mathrm{nl}$ lineshape on $\phi$ and $\zeta_{x'y'}$. (a) $R_\mathrm{nl}$ vs. $B_y$ at different $\phi$ when $\zeta_{x'y'}=20$. Curves are normalized to the value of $R_\mathrm{nl}$ when $\phi=90 \degree$. (b)$ \ R_\mathrm{nl}$ vs. $B_y$ for different $\zeta_{x'y'}$ when $\phi= 40 \degree$. $\zeta_{x'y'}$ is modulated by changing $\tau_{y'}$ and keeping $\tau_{x'}$ constant at 600$\ $ps. Curves are normalized to the value of $R_\mathrm{nl}$ when $\zeta_{x'y'}=1$. The values of all the other parameters are listed in Table$\ $\ref{Box:Par1}.}
 \label{fig:SV_dep}
\end{figure}

\begin{figure} [h]
\centering
\includegraphics[width=0.97\linewidth]{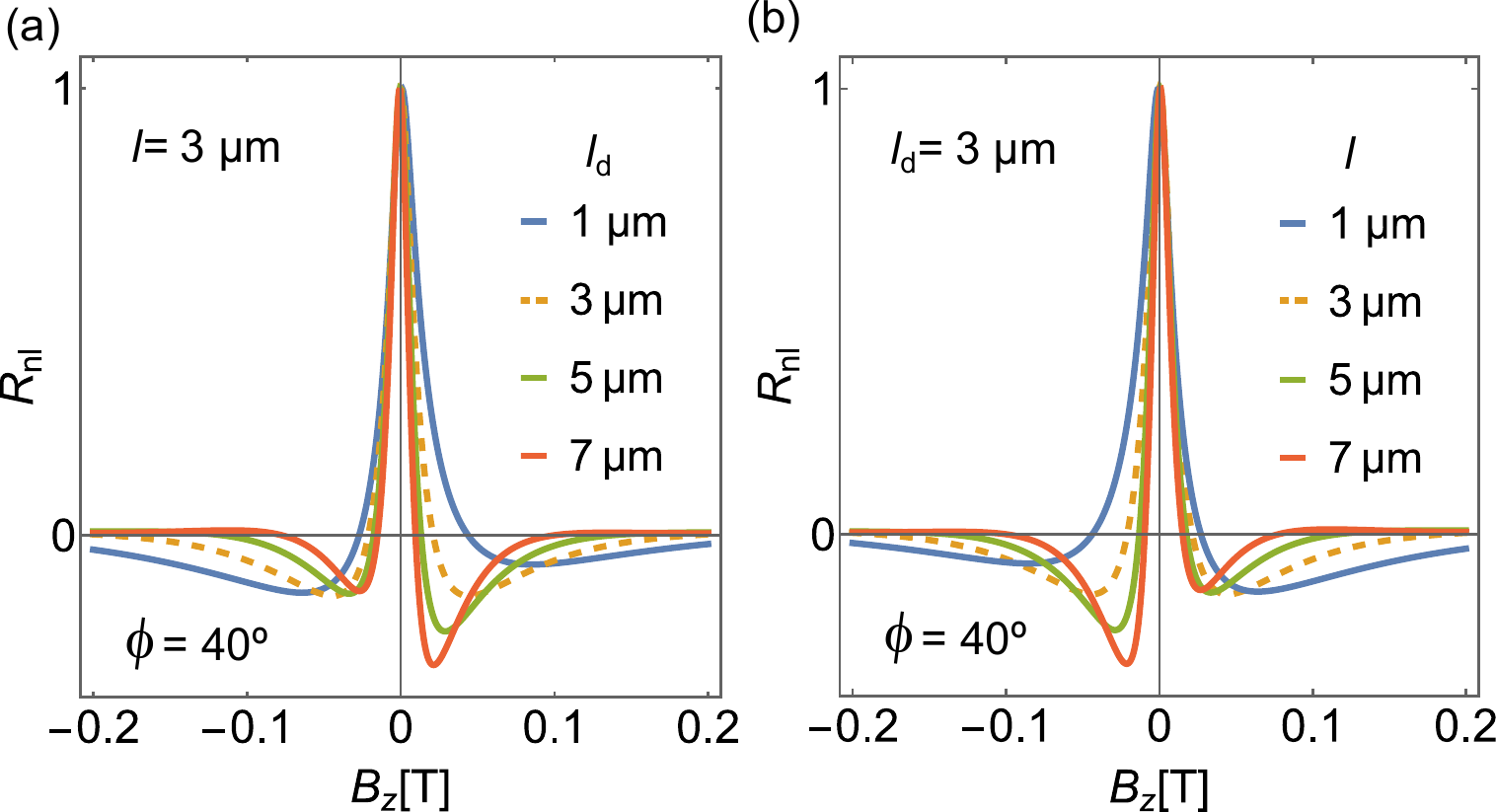}
\caption{Dependence of the in-plane spin precession lineshape on the device geometry. (a) $R_\mathrm{nl}$ vs. $B_z$ precession curves for different $l_\mathrm{d}$ and constant $l= 3 \, \micro$m. (b) $R_\mathrm{nl}$ vs. $B_z$ precession curves for different $l$ and constant $l_\mathrm{d}= 3 \, \micro$m. Notice that the curves in (a) are a mirror image of the curves in (b). All the curves are individually normalized to the value of $R_\mathrm{nl}$ at $B=0$. Spin transport parameters and $w_\mathrm{H}$ are listed in Table$\ $\ref{Box:Par1}.}
 \label{fig:HOP_Dist-dep}
\end{figure}

\begin{figure*} [!ht]
\centering
\includegraphics[width=0.85\linewidth]{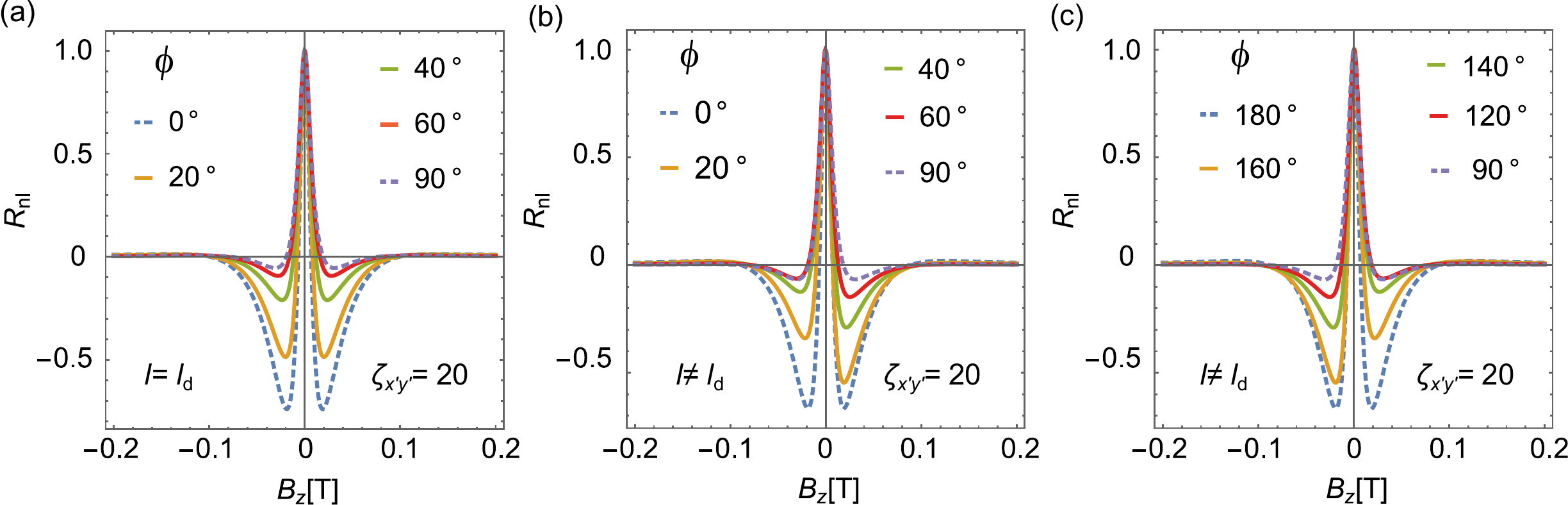}
\caption{Spin precession lineshape under an out-of-plane magnetic field $B_z$ as a function of  $\phi$. (a) $R_\mathrm{nl}$ vs. $B_z$ precession curves for selected $\phi$ in a symmetric device ($l=l_\mathrm{d}= 5 \ \micro$m). The curves remain symmetric about $B_z=0$ for all $\phi$. (b, c) $R_\mathrm{nl}$ vs. $B_z$ for different $\phi$ in an asymmetric device ($l=3 \ \micro$m, $ l_\mathrm{d}= 7 \ \micro$m) with $\phi \in [0\degree,90\degree]$ in (b) and $\phi \in [90\degree,180\degree]$ in (c). The curves in (b) are a mirror image of the curves in (c) with respect to $B_z=0$. Even in an asymmetric device, the curves are symmetric when $\phi=\pi/2 \ n$ (dashed lines). Spin transport parameters and $w_\mathrm{H}$ are listed in Table$\ $\ref{Box:Par1}.}
 \label{fig:HOP_Angle-dep}
\end{figure*}

\section{\label{sec:inplane}Anisotropic spin precession with in-plane magnetic fields}
We first consider the case where the magnetic field is applied in-plane, along the direction of the injected spins ($B_y$). For isotropic spin transport, this magnetic field geometry results in no spin precession, as the field is collinear with the spin polarization. However, as previously discussed, for anisotropic in-plane spin transport, spins are rotated toward $\hat{x}'$ within the anisotropic region, leading to precession around $B_y$. As shown in Fig.$\ $\ref{fig:SV_dep}(a), the magnitude of the spin signal increases as the angle $\phi$ increases from $0\degree$ to 90$\degree$. A decay in $R_\mathrm{nl}$ vs. $B_y$ is observed only for intermediate angles ($\phi \neq \pi/2 \ n$), \textit{i.e.}, when the spins are rotated away from $\hat{y}$. This anomalous decrease is caused by the spin precession in the anisotropic region around $B_y$, which reorients the spin polarization toward directions with shorter spin lifetimes ($\hat{y}'$ and $\hat{z}'$). Figure$ \, $\ref{fig:SV_dep}(b) illustrates how varying $\zeta_{x'y'}$ affects the lineshape of  $R_\mathrm{nl}$ vs. $B_y$ curves when $\phi= 40\degree$. Here, $\zeta_{x'y'}$ is tuned by changing $\tau_{y'}$ while keeping $\tau_{x'}$ fixed at 600 ps. For larger anisotropies, the decay is clearly visible because spins rotate substantially toward $\hat{x}'$, and a relatively large $\mu_x$ component of the spin density precesses when $B_y$ is applied. However, when $\zeta_{x'y'}$ is small, spin rotation away from $\hat{y}$ is minimal, and thus nearly no spin precession and decay occur when applying $B_y$. This makes the experimental observation of small anisotropic ratios challenging as the small decay can be hidden easily in the noise.

\section{Anisotropic spin precession with out-of-plane magnetic fields} \label{sec:l_ld_dep}

When the magnetic field is applied out of plane ($B_z$), the spins precess in the plane and the resulting spin dynamics is sensitive to $\tau_{x'}$ and $\tau_{y'}$ but does not depend on $\tau_z$. The spin precession curves for heterogeneous in-plane anisotropic devices exhibit novel features that are highly sensitive to $\zeta_{x'y'}$, $\phi$ and the positions of the injector and detector relative to the anisotropic region. In the case of a magnetic field applied along $B_y$, only the total injector-to-detector separation $L$ influences the signal, affecting its magnitude while leaving the lineshape unchanged. In contrast, the lineshape of $R_\mathrm{nl}$ vs. $B_z$ curves depends critically on the device design, namely the spin injector and detector distances from the proximitized region, $l$ and $l_\mathrm{d}$, respectively [see Fig.  \ref{fig:RegMod}(a) for definitions of $l$ and $l_\mathrm{d}$].

Figures$ \, $\ref{fig:HOP_Dist-dep}(a) and (b) show $R_\mathrm{nl}$ vs. $B_z$ curves for selected values of $l_\mathrm{d}$ and $l$. When $l_\mathrm{d} = l$, the side lobes, which correspond to a collective $\pi$ rotation of the spins, have the same amplitude and the spin precession curve is symmetric with respect to $B_z$. This symmetry breaks when $l_\mathrm{d}\neq l$. In such cases, the side lobes become asymmetric, with the deeper minimum appearing at positive (negative) $B_z$ when $l_\mathrm{d}>  l$ ($l_\mathrm{d}< l$).

Experimentally, this mirroring effect can be observed by swapping the injector and detector contacts in devices with an off-centered anisotropic region. Therefore, the distances of the injector/detector electrodes from the proximitized region are important design parameters when probing the in-plane spin anisotropy. Visualizing in-plane anisotropy becomes more challenging in symmetric devices with $l \approx l_\mathrm{d}$, as the spin precession curves remain roughly symmetric about $B_z= 0$. In such devices, the angle $\phi$ affects the depth of both side lobes equally [see Fig.$ \ $\ref{fig:HOP_Angle-dep}(a)], and the anisotropy can be identified only by close examination of the overall spin precession profile (\textit{e.g.}, by comparing the magnitude of the lobes relative to the spin signal at $B_z=0$). In contrast, asymmetric devices exhibit clearly unequal lobe depths for all $\phi$, except at angles $\pi/2 \ n$ [Fig.$ \ $\ref{fig:HOP_Angle-dep}(b) and (c)]. Consequently, designing asymmetric devices offers a practical advantage for experimentally detecting in-plane spin anisotropy.
\begin{figure} [!ht]
\centering
\includegraphics[width=0.99\linewidth]{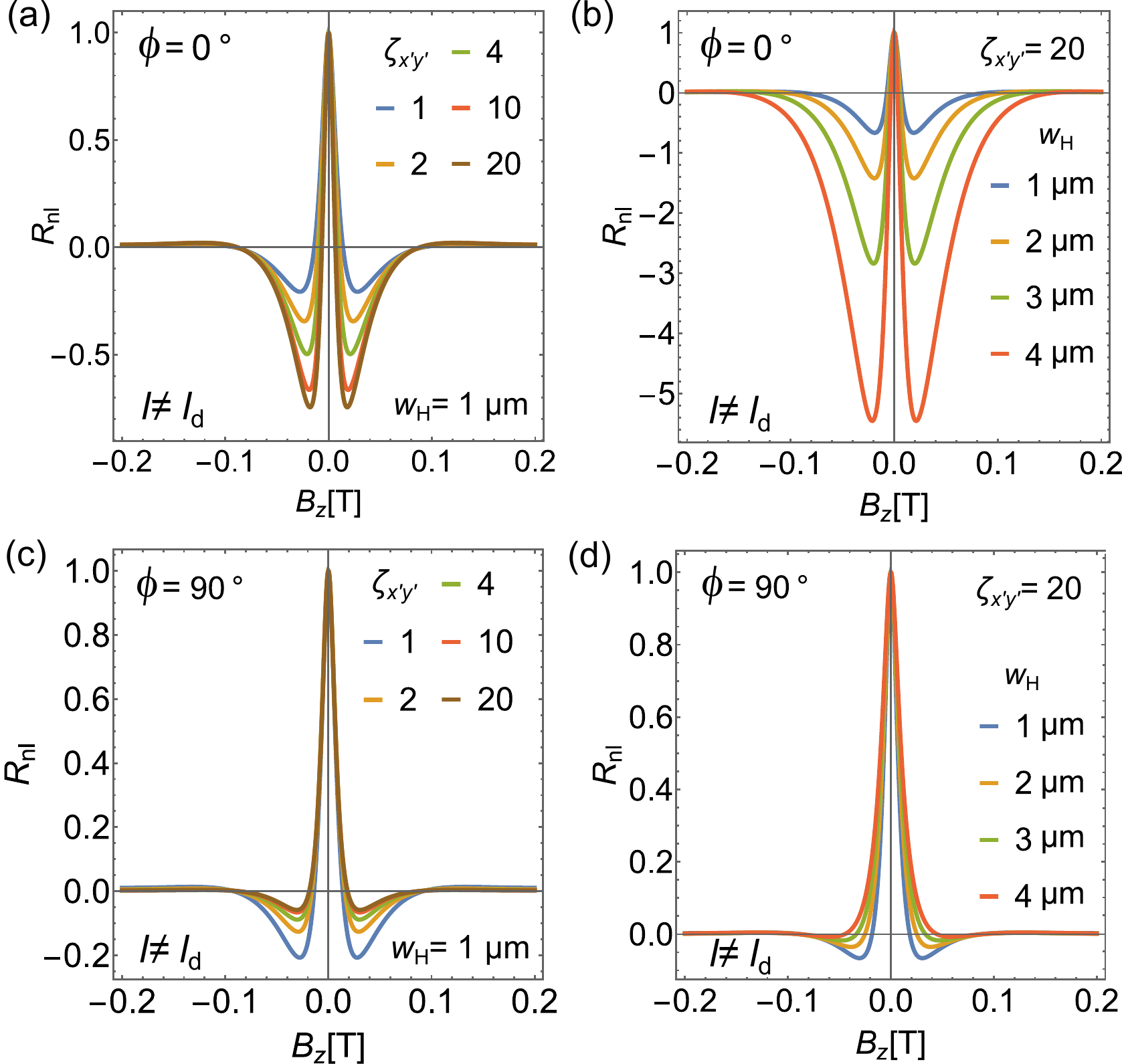}
\caption{Dependence of the in-plane precession lineshape on $\zeta_{x'y'}$ [(a) and (c)] and $w_\mathrm{H}$ [(b) and (d)] for $\phi= 0\degree$ [(a) and (b)] and $\phi= 90\degree$ [(c) and (d)]. All the other parameters are listed in Table$\ $\ref{Box:Par1}.}
 \label{fig:HOP_Thickness_dep}
\end{figure}

Visualizing the anisotropy can be somewhat challenging in asymmetric devices for small $\zeta_{x'y'}$ or $w_\mathrm{H}$ when $\phi =\pi/2 \ n$, as shown in Figure$ \ $\ref{fig:HOP_Thickness_dep} for $\phi= 0 \degree$ [Fig.$\ $ \ref{fig:HOP_Thickness_dep} (a) and (b)] and $\phi= 90 \degree$ [Fig.$\ $\ref{fig:HOP_Thickness_dep}(c) and (d)]. The precession curves at $\phi=0 \degree$ resemble the out-of-plane precession behavior ($R_\mathrm{nl}$ vs. $B_x$) observed in graphene/2H-TMDC heterostructures, where the longest spin lifetime aligns with $\hat{z}$ \citep{ghiasi_large_2017, benitez_strongly_2018}. In contrast, when $\phi \approx 90\degree$, the spin anisotropy becomes more difficult to detect, as the lineshape resembles the isotropic case. In this scenario, increasing the width of the anisotropic region enhances the visibility of the spin anisotropy, eventually leading to complete suppression of the negative oscillation [Fig.$ \ $\ref{fig:HOP_Thickness_dep}(d)].

\section{Anisotropic spin precession under oblique magnetic fields} \label{sec:simul-obl}

 \begin{figure} [h]
\centering
\includegraphics[width=0.99\linewidth]{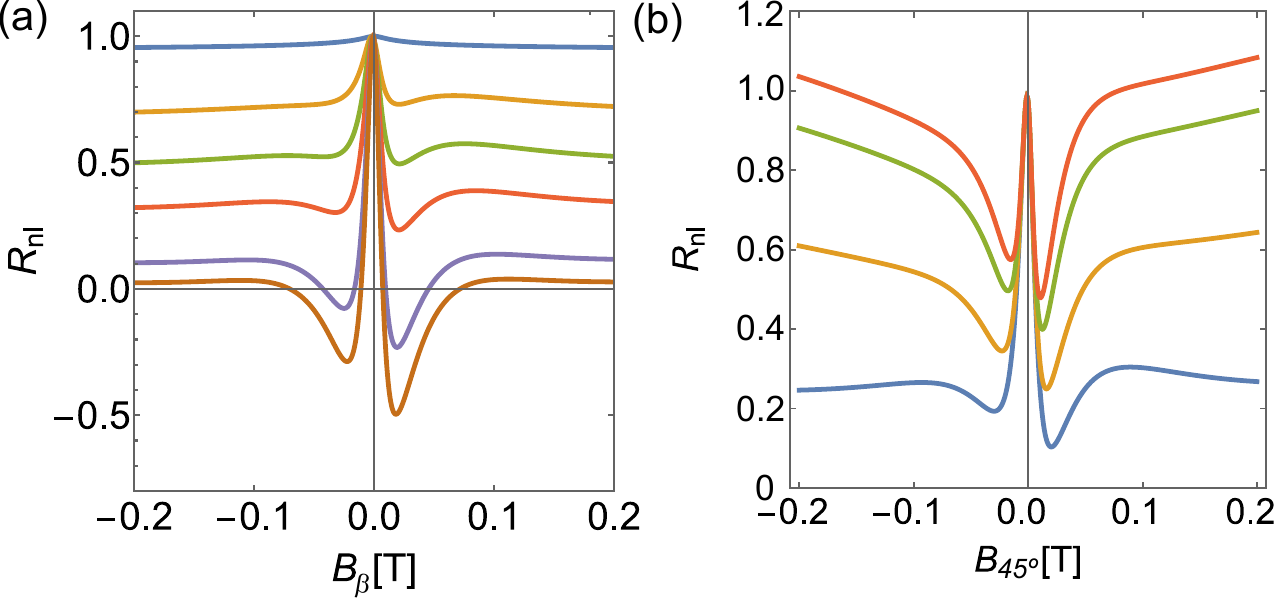}
\caption{(a) Oblique-field precession curves for an asymmetric device with $\phi= 10 \degree$ and $\beta= 0\degree,20\degree,30\degree,40\degree,60\degree,90\degree$ (from top to bottom). (b) Dependence of the spin precession lineshape on $\tau_z$ at $\beta= 45\degree$ with $\phi= 10 \degree$. Curves correspond to $\tau_z= 400,100,25, \mathrm{and} \ 5 \ $ps (from top to bottom). Because for $\beta= 45\degree$ spins can acquire a full out-of-plane orientation during precession, this geometry is particularly sensitive to $\tau_z$. }
 \label{fig:Obliqu_betaDep}
\end{figure}

Out-of-plane spin precession can be induced by applying an in-plane magnetic field perpendicular to the easy axis of the FM injector and detector ($B_x$). However, when performing these experiments, the small saturation field of the FMs along $\hat{x}$ complicates quantitative data analysis and the extraction of an accurate value of the spin anisotropy \cite{raes_spin_2017}. A more precise determination of the  out-of-plane spin lifetime $\tau_z$ can be achieved through oblique spin precession measurements. When an oblique magnetic field is applied—characterized by the angle $\beta$—the injected spins precess around this tilted field, acquiring an out-of-plane spin component. As a result, spin dynamics is sensitive to $\tau_z$ in addition to $\tau_{x'}$ and $\tau_{y'}$ \cite{raes_determination_2016}. Figure$ \, $\ref{fig:Obliqu_betaDep}{(a)} shows the simulated precession curves for an asymmetric device ($l=1 \ \micro$m, $l_\mathrm{d}= 9 \ \micro$m) at representative oblique angles $\beta$, when $\tau_{x'}= 200 \ \mathrm{ps}, \tau_{y'}= 20 \ \mathrm{ps},\tau_{z}= 5 \ \mathrm{ps}$, and $\phi=10\degree$. We can observe that due to in-plane anisotropy, the decrease as a function of the magnetic field magnitude, described in Fig.  \ref{fig:SV_dep} is observed for $\beta \neq 90$. In practice, the experimental in-plane anisotropy can be extracted using measurements with an applied out-of-plane magnetic field, as discussed in Section V. Then, having estimations of $\tau_{x'}$ and $\tau_{y'}$, one can fit the oblique spin precession results as those in Fig.$ \, $\ref{fig:Obliqu_betaDep}{(a)} to obtain $\tau_{z'}$.
Figure$ \, $\ref{fig:Obliqu_betaDep}(b) shows pronounced changes in the lineshape of the precession curves at $\beta=45\degree$ (\textit{i.e}, angle at which spin polarization can be brought completely out-of-plane) when $\phi= 10 \degree$ and $\tau_z$ is gradually reduced from 400 ps to 5 ps. This shows the high sensitivity of the oblique geometry to the value of $\tau_{z}$.

\section{Experimental results}\label{sec:exp}
In the previous sections, we implemented the solution of the spin-diffusion model for a heterogeneous nonlocal spin device. We discussed how the local spin-dependent electrochemical potential evolves with the anisotropy ratios and device geometry, including the orientation of the principal axes of spin relaxation. We also related the evolution of the spin-dependent electrochemical potential as a function of the magnetic field $B$ to measurable quantities, such as the nonlocal resistance $R_\mathrm{nl}$. Here, we compare experimental and calculated $R_\mathrm{nl}$ for relevant device configurations and detail the process used to extract anisotropy parameters from the measurements.  The experimental results correspond to a lateral graphene spin device in which the spin channel is partially covered by PdSe$_2$. This low-symmetry material has a puckered pentagonal structure and displays pronounced anisotropic optical, electrical, and thermal properties \citep{yu_direct_2020,lu_layer-dependent_2020,xu_unravelling_2024}. When thinned down, PdSe$_2$ becomes semiconducting \citep{sun_electronic_2015,oyedele_pdse2_2017} and thus constitutes a perfect candidate for investigating novel proximity-induced SOC in graphene, as parallel conduction and spin absorption in PdSe$_2$ can be suppressed by tuning the Fermi level into the bandgap through gating.

The device was fabricated on \textit{p}-doped silicon substrate covered by thermal oxide (440 nm), which can be used as an effective back-gate ($V_\mathrm{g}$) to tune the graphene's carrier density. The flakes were obtained by mechanically exfoliating graphene from a highly oriented pyrolytic graphite source (SPI Supplies) and a bulk crystal of PdSe$_2$ grown by the self-flux method using selenium as flux. We selected the graphene flake to be approximately 1 $\micro$m wide and several tens of $\micro$m long, and a thin PdSe$_2$ flake with a width of about 1 $\micro$m. The PdSe$_2$ flake was transferred onto graphene (see details in Ref. \citep{sierra_room-temperature_2025}). The stack was then annealed at 280$\, \degree$C in high vacuum to improve the quality of the interface between graphene and PdSe$_2$. Two types of contacts were defined using electron-beam lithography. In the first step, we patterned outer metallic contacts made of Ti/Pd (2 nm/25 nm). This was followed by a second step of e-beam lithography to define the inner contacts, made of TiO$_x$/Co (1$\, $nm/30$ \ $nm), which act as spin injectors and detectors. To enhance the contrast due to in-plane anisotropy,  injector and detector were designed to have different distances from the graphene/PdSe$_2$ bilayer region  (see Section \ref{sec:l_ld_dep}).

\begin{figure} [hb]
\centering
\includegraphics[width=0.99\linewidth]{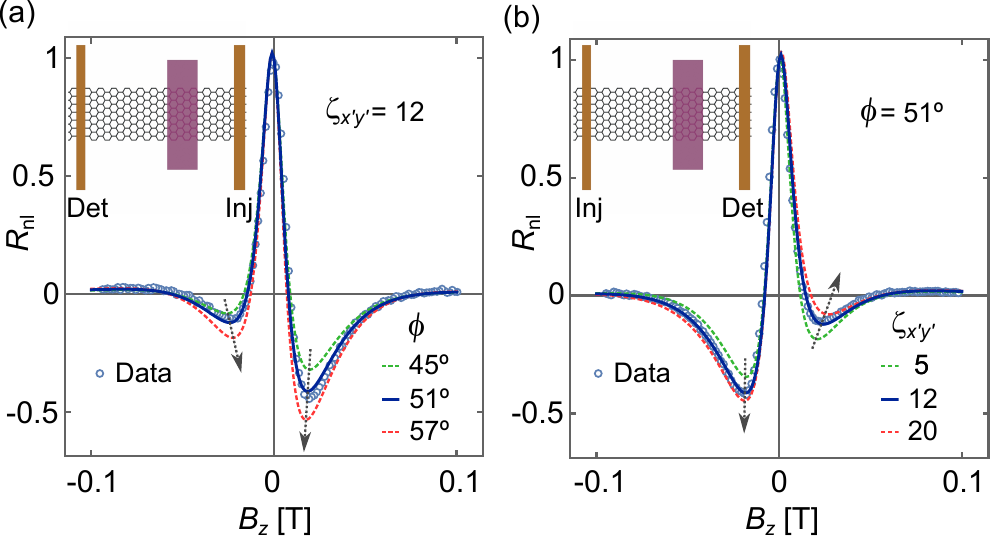}
\caption{Experimental $R_\mathrm{nl}$ (blue circles) and modeled precession curves (lines) as a function of $B_z$. Solid blue lines are the best-fit curves with $ \phi=51 \degree$ and anisotropy $ \zeta_{x'y'}= 12$. Dotted curves show calculations with parameters deviating from the best-fit values. Gray arrows indicate the direction in which the minima shift as $ \phi$ increases in (a) and as $ \zeta_{x'y'}$ increases in (b). (a) Configuration in which spin injector is closer to the anisotropic region than the detector. (b) Configuration in which the spin injector is further away from the anisotropic region than the detector.}
 \label{fig:DataHOP}
\end{figure}

Figure \ref{fig:DataHOP} shows the experimental data measured at $V_\mathrm{g} = -30 \ $V together with the modeled lineshape of the spin precession curves obtained when $B$ is applied out of plane. Measurements were performed in two configurations in which the injector and detector are swapped. The data acquired in these two configurations are mirror images of each other about $B_z = 0$, in agreement with the prediction of the model (compare with Fig. \ref{fig:HOP_Dist-dep}). The different depths of the two side lobes indicate the presence of a persistent in-plane SOF oriented along a direction different from $\phi =\pi/2 \ n$ [see Figs. \ref{fig:HOP_Angle-dep}(b) and (c)]. Moreover, the large difference in the magnitudes of the side lobes indicates a pronounced in-plane spin relaxation anisotropy between spins oriented along $\hat{x}'$ and $\hat{y}'$. Indeed, we find that $\phi= 51\degree$ and $\zeta_{x'y'}= 12$ provide the best fit to the experimental data (solid dark lines). Figure \ref{fig:DataHOP}(a) further illustrates the evolution of the lineshape when the anisotropy is fixed at $\zeta_{x'y'}= 12$ but the angle $\phi$ is changed to $45\degree$ and $57\degree$ (dashed green and red lines, respectively). A few-degree deviation in the SOF direction results in noticeable discrepancies between the modeled and measured lineshapes. Similarly, Fig. \ref{fig:DataHOP}(b) shows the variation in the shape of the lines when $\phi$ is fixed at 51$\degree$ and the in-plane spin anisotropy parameter $\zeta_{x'y'}$ is decreased or increased to 5 and 20 (dashed green and red lines, respectively). Although the difference appears less pronounced in this case, particularly for larger anisotropic ratios, here we analyze only the normalized lineshape. In practice, variations in $\tau_{x'y'}$ lead to substantial changes in the absolute magnitude of the measured spin signals, which restrict the range of possible spin lifetimes. Furthermore, while a change in $\phi$ results in variations in the magnitude of the minima at positive and negative $B$ in the same direction, changes in $\zeta_{x'y'}$ cause an increase of one minimum and decrease of the other (see the gray dotted arrows in Fig.$\ $\ref{fig:DataHOP}). This indicates that the individual effects of $\phi$ and $\zeta_{x'y'}$ are not correlated and it is possible to obtain reliable values by fitting.

\begin{figure} [!ht]
\centering
\includegraphics[width=0.99\linewidth]{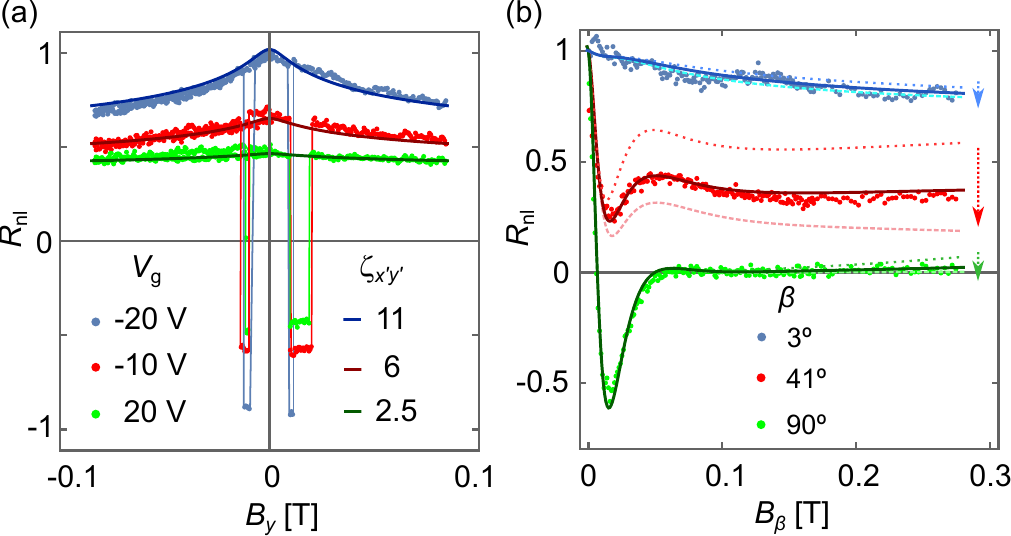}
\caption{Experimental $R_\mathrm{nl}$ (solid circles) and modeled lineshape (solid lines) as a function of $B_y$  for different $V_\mathrm{g}$ (a) and as a function of $B_\beta$ for selected $\beta$ (b). (a) The model shows that sweeping $V_\mathrm{g}$ from -20 V to 20 V modulates $\zeta_{x'y'}$ from 11 to 2.5. (b) $R_\mathrm{nl}$ vs $B_{\beta}$ for $\beta= 3\degree, 41\degree, 90\degree$ at $V_\mathrm{g}= 5 \ $V. The best fits (solid lines) are obtained with $\tau_z =15$ ps. Changing $\tau_z$ to 30$\ $ps or 6$\ $ps leads to significant deviations in the oblique configuration $\beta= 41\degree$ (dotted and dashed red lines).}
 \label{fig:DataSV-Obl}
\end{figure}

We now focus on alternative magnetic field configurations. Figure \ref{fig:DataSV-Obl}(a) shows the spin signal at various $V_\mathrm{g}$ values when the magnetic field is applied along the easy axis of the FM injector and detector contacts. In this configuration, the aggregate spins effectively rotate upon reaching the proximitized region because of the presence of an in-plane SOF. Since, in the general case, this SOF is not aligned with the direction of the incoming spins, spin components misaligned with it relax faster within the proximitized region, effectively rotating towards $\hat{x}'$, \textit{i.e}., the in-plane axis associated with the longest spin relaxation time. As a result, the spins are no longer parallel to the applied magnetic field (their initial alignment) and start to precess, acquiring an out-of-plane spin component. This leads to a reduction in the measured spin signal as the magnetic field increases, with a decay rate that strongly depends on the in-plane anisotropy ratio $\zeta_{x'y'}$ and, to less extent, $\tau_z$ [compare with Fig.  \ref{fig:SV_dep}]. Figure$\ $\ref{fig:DataSV-Obl}(b) presents the experimental results obtained when $B$ is applied at representative oblique angles $\beta$, together with the corresponding modeled precession curves (lines). Importantly, these measurements rule out magnetization canting as the origin of the field-asymmetry observed in Fig. \ref{fig:DataHOP}, since the observed suppression of the signal is incompatible with a progressive correction of any magnetization misalignment, which would be expected as the magnetic field increases.

The close match between experimental data and simulation when $\tau_z= 15$ ps (solid lines) highlights the high accuracy of the model in reproducing the experimental results. The dashed and dotted lines demonstrate the sensitivity of the signal to variations in $\tau_z$. It is observed that, when $B$ is out of plane, \textit{i.e.}, $\beta$=$90\degree$ (green dots and lines), the change is very small and is due to the tilting of the magnetization of the electrodes (which is considered in the model). In addition, the general normalized profile when $B$ is aligned with the injected spins also changes weakly with $\tau_z$ (blue dots and lines). This is because the out-of-plane spin density is largest away from the anisotropic region but stays close to zero within [see the dotted blue line in Fig.$\ $\ref{fig:RegMod}(b)]. Only when the spins are forced out of plane at the location of the anisotropic region, \textit{i.e.}, when $\beta$ approaches $45\degree$, we observe high sensitivity to $\tau_z$ (red lines).

\section{Conclusions}
We have presented a comprehensive solution of the Bloch diffusion equation for heterogeneous spin devices. By calculating the spatial distribution of spin density and simulating the precession lineshapes under various geometries, spin transport parameters, and magnetic field orientations, we have identified clear signatures of spin anisotropy. A key finding is that misalignment between the spin injection direction and the principal spin-lifetime axes in the anisotropic region induces an effective rotation of the spin accumulation that propagates into adjacent regions. This effect manifests experimentally as an anomalous reduction of the nonlocal resistance
$R_\mathrm{nl}$ when the magnetic field is applied parallel to the injected spin direction—a direct indication of in-plane anisotropy. This phenomenon originates from spin precession of the rotated spins, which forces spins toward the short-lived axes. Another distinct fingerprint of in-plane anisotropy is the observed asymmetry between positive and negative magnetic fields in out-of-plane and oblique precession curves. Such asymmetries arise in devices in which the anisotropic region is positioned asymmetrically with respect to the injector and detector. Conversely, symmetry is restored when the spin injection axis aligns with $x'$ or $y'$, or when the anisotropic region is equidistant from the injector and detector. In these cases, especially when anisotropy is weak, or spin transport parameters in isotropic regions are not precisely known, the spin precession curves may appear isotropic upon quick inspection, and a lineshape analysis based on the full solution of the diffusion equation is required for a reliable characterization. Based on these insights, we conclude that asymmetric devices in which the easy magnetization axes of ferromagnetic electrodes are deliberately misaligned with the principal spin-lifetime axes of the anisotropic region offer the most sensitive configuration for detecting and quantifying in-plane spin anisotropy, enabling robust extraction of primary spin-relaxation axes and anisotropy parameters. The anisotropy ratios are best quantified by first using out-of-plane magnetic fields together with measurements with the field aligned with the injector magnetization to determine the in-plane anisotropy, followed by oblique spin precession at $\beta \approx 45\degree$ and other intermediate oblique angles to extract the out-of-plane spin lifetime.

These results provide a practical roadmap for symmetry-resolved characterization of spin relaxation in heterogeneous 2D devices under realistic measurement geometries, including design rules for tailoring device layout and field orientation to maximize sensitivity and to benchmark proximity-induced SOC in graphene-based heterostructures. The approach is readily transferable to other material combinations and crystalline symmetries, resulting in a unified methodology to disentangle in-plane and out-of-plane lifetimes and to identify the principal relaxation axes and spin-orbit fields.

\section{Acknowledgments}
This research was partially supported by MICIU/AEI/10.13039/501100011033 through grants no. PID2022-143162OB-I00 and Severo Ochoa Programme CEX2021-001214-S, FEDER. J.S. thanks H2020 Marie Skłodowska-Curie grant agreement no. 754558. D.D. and V. M. acknowledge support of the Bulgarian National Science Fund under Contract No. KP-06-N-58/12. 

\section{Data Availability}
The data that support the experimental findings of this article are openly available at \cite{Zenodo_2025}.

\bibliography{references}

@article{Zenodo_2025,
url = {https://doi.org/10.5281/zenodo.17484542},
year = {2025},
journal = {Zenodo},
author = {J. Sv\v{e}tlík and J. F. Sierra and L. Camosi and  W. Savero Torres and F.
Herling and V. Marinova and D. Dimitrov and S. O. Valenzuela},
title = {Experimental data for "{H}anle Lineshapes and Spin-Rotation..." [{D}ata set]}
}

@article{benitez_tunable_2020,
	title = {Tunable room-temperature spin galvanic and spin {Hall} effects in van der {Waals} heterostructures},
	volume = {19},
	copyright = {2020 The Author(s), under exclusive licence to Springer Nature Limited},
	issn = {1476-4660},
	url = {https://www.nature.com/articles/s41563-019-0575-1},
	doi = {10.1038/s41563-019-0575-1},
	number = {2},
	urldate = {2020-06-12},
	journal = {Nature Materials},
	author = {Benítez, L. Antonio and Savero Torres, Williams and Sierra, Juan F. and Timmermans, Matias and Garcia, Jose H. and Roche, Stephan and Costache, Marius V. and Valenzuela, Sergio O.},
	month = feb,
	year = {2020},
	pages = {170--175},
}

@article{drogeler_2016,
URL = {https://doi.org/10.1021/acs.nanolett.6b00497},
author = {Dr{\"o}geler, Marc and Franzen, Christopher and Volmer, Frank and Pohlmann, Tobias and Banszerus, Luca and Wolter, Maik and Watanabe, Kenji and Taniguchi, Takashi and Stampfer, Christoph and Beschoten, Bernd},
title = {Spin Lifetimes Exceeding 12 ns in Graphene Nonlocal Spin Valve Devices},
journal = {Nano Letters},
volume = {16},
number = {6},
pages = {3533-3539},
year = {2016}
}

@article{Gebeyehu_2019,
doi = {10.1088/2053-1583/ab1874},
url = {https://doi.org/10.1088/2053-1583/ab1874},
year = {2019},
month = {may},
publisher = {IOP Publishing},
volume = {6},
number = {3},
pages = {034003},
author = {Gebeyehu, Z M and Parui, S and Sierra, J F and Timmermans, M and Esplandiu, M J and Brems, S and Huyghebaert, C and Garello, K and Costache, M V and Valenzuela, S O},
title = {Spin communication over 30 µm long channels of chemical vapor deposited graphene on SiO2},
journal = {2D Materials}
}

@article{cummings_giant_2017,
	title = {Giant {Spin} {Lifetime} {Anisotropy} in {Graphene} {Induced} by {Proximity} {Effects}},
	volume = {119},
	url = {https://link.aps.org/doi/10.1103/PhysRevLett.119.206601},
	doi = {10.1103/PhysRevLett.119.206601},
	number = {20},
	urldate = {2020-07-13},
	journal = {Phys. Rev. Lett.},
	author = {Cummings, Aron W. and Garcia, Jose H. and Fabian, Jaroslav and Roche, Stephan},
	month = nov,
	year = {2017},
	pages = {206601},
}

@article{valenzuela_nonlocal_2009,
	title = {Nonlocal electronic spin detection, spin accumulation and the spin {Hall} effect},
	volume = {23},
	issn = {0217-9792},
	url = {https://www.worldscientific.com/doi/abs/10.1142/S021797920905290X},
	doi = {10.1142/S021797920905290X},
	number = {11},
	urldate = {2021-04-14},
	journal = {Int. J. Mod. Phys. B},
	author = {Valenzuela, Sergio O.},
	month = apr,
	year = {2009},
	pages = {2413--2438},
}

@article{gmitra_graphene_2015,
	title = {Graphene on transition-metal dichalcogenides: {A} platform for proximity spin-orbit physics and optospintronics},
	volume = {92},
	shorttitle = {Graphene on transition-metal dichalcogenides},
	url = {https://link.aps.org/doi/10.1103/PhysRevB.92.155403},
	doi = {10.1103/PhysRevB.92.155403},
	number = {15},
	urldate = {2021-05-04},
	journal = {Phys. Rev. B},
	author = {Gmitra, Martin and Fabian, Jaroslav},
	month = oct,
	year = {2015},
	pages = {155403},
}

@article{herling_gate_2020,
	title = {Gate tunability of highly efficient spin-to-charge conversion by spin {Hall} effect in graphene proximitized with {WSe$_2$}},
	volume = {8},
	url = {https://aip.scitation.org/doi/10.1063/5.0006101},
	doi = {10.1063/5.0006101},
	number = {7},
	urldate = {2021-06-15},
	journal = {APL Materials},
	author = {Herling, Franz and Safeer, C. K. and Ingla-Aynés, Josep and Ontoso, Nerea and Hueso, Luis E. and Casanova, Fèlix},
	month = jul,
	year = {2020},
	pages = {071103},
}

@article{han_graphene_2014,
	title = {Graphene spintronics},
	volume = {9},
	copyright = {2014 Nature Publishing Group, a division of Macmillan Publishers Limited. All Rights Reserved.},
	issn = {1748-3395},
	url = {https://www.nature.com/articles/nnano.2014.214},
	doi = {10.1038/nnano.2014.214},
	number = {10},
	urldate = {2021-12-09},
	journal = {Nature Nanotech},
	author = {Han, Wei and Kawakami, Roland K. and Gmitra, Martin and Fabian, Jaroslav},
	month = oct,
	year = {2014},
	pages = {794--807},
}

@article{johnson_interfacial_1985,
	title = {Interfacial charge-spin coupling: {Injection} and detection of spin magnetization in metals},
	volume = {55},
	shorttitle = {Interfacial charge-spin coupling},
	url = {https://link.aps.org/doi/10.1103/PhysRevLett.55.1790},
	doi = {10.1103/PhysRevLett.55.1790},
	number = {17},
	urldate = {2024-06-20},
	journal = {Phys. Rev. Lett.},
	author = {Johnson, Mark and Silsbee, R. H.},
	month = oct,
	year = {1985},
	pages = {1790--1793},
}

@article{sierra_van_2021,
	title = {Van der {Waals} heterostructures for spintronics and opto-spintronics},
	volume = {16},
	copyright = {2021 Springer Nature Limited},
	issn = {1748-3395},
	url = {https://www.nature.com/articles/s41565-021-00936-x},
	doi = {10.1038/s41565-021-00936-x},
	number = {8},
	urldate = {2024-08-29},
	journal = {Nat. Nanotechnol.},
	author = {Sierra, Juan F. and Fabian, Jaroslav and Kawakami, Roland K. and Roche, Stephan and Valenzuela, Sergio O.},
	month = aug,
	year = {2021},
	pages = {856--868},
}

@article{torrey_bloch_1956,
	title = {Bloch {Equations} with {Diffusion} {Terms}},
	volume = {104},
	url = {https://link.aps.org/doi/10.1103/PhysRev.104.563},
	doi = {10.1103/PhysRev.104.563},
	number = {3},
	urldate = {2024-08-29},
	journal = {Phys. Rev.},
	author = {Torrey, H. C.},
	month = nov,
	year = {1956},
	pages = {563--565},
}

@article{raes_determination_2016,
	title = {Determination of the spin-lifetime anisotropy in graphene using oblique spin precession},
	volume = {7},
	copyright = {2016 The Author(s)},
	issn = {2041-1723},
	url = {https://www.nature.com/articles/ncomms11444},
	doi = {10.1038/ncomms11444},
	number = {1},
	urldate = {2024-08-29},
	journal = {Nat Commun},
	author = {Raes, Bart and Scheerder, Jeroen E. and Costache, Marius V. and Bonell, Frédéric and Sierra, Juan F. and Cuppens, Jo and Van de Vondel, Joris and Valenzuela, Sergio O.},
	month = may,
	year = {2016},
	pages = {11444},
}

@article{ghiasi_large_2017,
	title = {Large {Proximity}-{Induced} {Spin} {Lifetime} {Anisotropy} in {Transition}-{Metal} {Dichalcogenide}/{Graphene} {Heterostructures}},
	volume = {17},
	issn = {1530-6984},
	url = {https://doi.org/10.1021/acs.nanolett.7b03460},
	doi = {10.1021/acs.nanolett.7b03460},
	number = {12},
	urldate = {2024-10-03},
	journal = {Nano Lett.},
	author = {Ghiasi, Talieh S. and Ingla-Aynés, Josep and Kaverzin, Alexey A. and van Wees, Bart J.},
	month = dec,
	year = {2017},
	pages = {7528--7532},
}

@article{benitez_strongly_2018,
	title = {Strongly anisotropic spin relaxation in graphene–transition metal dichalcogenide heterostructures at room temperature},
	volume = {14},
	copyright = {2017 The Author(s)},
	issn = {1745-2481},
	url = {https://www.nature.com/articles/s41567-017-0019-2},
	doi = {10.1038/s41567-017-0019-2},
	number = {3},
	urldate = {2024-10-03},
	journal = {Nature Phys},
	author = {Benítez, L. Antonio and Sierra, Juan F. and Savero Torres, Williams and Arrighi, Aloïs and Bonell, Frédéric and Costache, Marius V. and Valenzuela, Sergio O.},
	month = mar,
	year = {2018},
	pages = {303--308},
}

@article{raes_spin_2017,
	title = {Spin precession in anisotropic media},
	volume = {95},
	url = {https://link.aps.org/doi/10.1103/PhysRevB.95.085403},
	doi = {10.1103/PhysRevB.95.085403},
	number = {8},
	urldate = {2024-10-03},
	journal = {Phys. Rev. B},
	author = {Raes, B. and Cummings, A. W. and Bonell, F. and Costache, M. V. and Sierra, J. F. and Roche, S. and Valenzuela, S. O.},
	month = feb,
	year = {2017},
	pages = {085403},
}

@article{sierra_room-temperature_2025,
	title = {Room-temperature anisotropic in-plane spin dynamics in graphene induced by {PdSe$_2$} proximity},
	volume = {24},
	copyright = {2025 The Author(s), under exclusive licence to Springer Nature Limited},
	issn = {1476-4660},
	url = {https://www.nature.com/articles/s41563-024-02109-2},
	doi = {10.1038/s41563-024-02109-2},
	number = {6},
	urldate = {2025-10-20},
	journal = {Nat. Mater.},
	author = {Sierra, Juan F. and Světlík, Josef and Savero Torres, Williams and Camosi, Lorenzo and Herling, Franz and Guillet, Thomas and Xu, Kai and Reparaz, Juan Sebastián and Marinova, Vera and Dimitrov, Dimitre and Valenzuela, Sergio O.},
	month = jun,
	year = {2025},
	pages = {876--882},
}

@incollection{valenzuela_spintronic_2024,
	address = {Oxford},
	title = {Spintronic materials},
	isbn = {978-0-323-91408-6},
	url = {https://www.sciencedirect.com/science/article/pii/B9780323908009002298},
	urldate = {2025-07-29},
	booktitle = {Encyclopedia of {Condensed} {Matter} {Physics} ({Second} {Edition})},
	publisher = {Academic Press},
	author = {Valenzuela, Sergio O. and Gambardella, Pietro and Garello, Kevin and Klein, Olivier and Sierra, Juan F. and Sinova, Jairo},
	editor = {Chakraborty, Tapash},
	month = jan,
	year = {2024},
	pages = {159--176},
}

@article{zollner_first-principles_2025,
	title = {First-principles determination of spin–orbit coupling parameters in two-dimensional materials},
	volume = {7},
	copyright = {2025 Springer Nature Limited},
	issn = {2522-5820},
	url = {https://www.nature.com/articles/s42254-025-00818-4},
	doi = {10.1038/s42254-025-00818-4},
	number = {5},
	urldate = {2025-07-29},
	journal = {Nat Rev Phys},
	author = {Zollner, Klaus and Kurpas, Marcin and Gmitra, Martin and Fabian, Jaroslav},
	month = may,
	year = {2025},
	pages = {255--269},
}

@article{li_oblique_2008,
	title = {Oblique {Hanle} effect in semiconductor spin transport devices},
	volume = {92},
	issn = {0003-6951},
	url = {https://doi.org/10.1063/1.2907497},
	doi = {10.1063/1.2907497},
	number = {14},
	urldate = {2025-07-29},
	journal = {Applied Physics Letters},
	author = {Li, Jing and Huang, Biqin and Appelbaum, Ian},
	month = apr,
	year = {2008},
	pages = {142507},
}

@article{motsnyi_optical_2003,
	title = {Optical investigation of electrical spin injection into semiconductors},
	volume = {68},
	url = {https://link.aps.org/doi/10.1103/PhysRevB.68.245319},
	doi = {10.1103/PhysRevB.68.245319},
	number = {24},
	urldate = {2025-07-29},
	journal = {Phys. Rev. B},
	author = {Motsnyi, V. F. and Van Dorpe, P. and Van Roy, W. and Goovaerts, E. and Safarov, V. I. and Borghs, G. and De Boeck, J.},
	month = dec,
	year = {2003},
	pages = {245319},
}

@article{hoque_spin-valley_2023,
	title = {Spin-valley coupling and spin-relaxation anisotropy in all-{CVD} {Graphene}-{MoS$_2$} van der {Waals} heterostructure},
	volume = {7},
	url = {https://link.aps.org/doi/10.1103/PhysRevMaterials.7.044005},
	doi = {10.1103/PhysRevMaterials.7.044005},
	number = {4},
	urldate = {2025-07-29},
	journal = {Phys. Rev. Mater.},
	author = {Hoque, Anamul Md. and Ramachandra, Vasudev and George, Antony and Najafidehaghani, Emad and Gan, Ziyang and Mitra, Richa and Zhao, Bing and Khokhriakov, Dmitrii and Turchanin, Andrey and Lara-Avila, Samuel and Kubatkin, Sergey and Dash, Saroj P.},
	month = apr,
	year = {2023},
	pages = {044005},
}

@article{garcia_spin_2018,
	title = {Spin transport in graphene/transition metal dichalcogenide heterostructures},
	volume = {47},
	issn = {1460-4744},
	url = {https://pubs.rsc.org/en/content/articlelanding/2018/cs/c7cs00864c},
	doi = {10.1039/C7CS00864C},
	number = {9},
	urldate = {2025-07-29},
	journal = {Chem. Soc. Rev.},
	author = {Garcia, Jose H. and Vila, Marc and Cummings, Aron W. and Roche, Stephan},
	month = may,
	year = {2018},
	pages = {3359--3379},
}

@article{jedema_electrical_2002,
	title = {Electrical detection of spin precession in a metallic mesoscopic spin valve},
	volume = {416},
	rights = {http://www.springer.com/tdm},
	issn = {0028-0836, 1476-4687},
	url = {https://www.nature.com/articles/416713a},
	doi = {10.1038/416713a},
	pages = {713--716},
	number = {6882},
	journal = {Nature},
	shortjournal = {Nature},
	author = {Jedema, F. J. and Heersche, H. B. and Filip, A. T. and Baselmans, J. J. A. and Van Wees, B. J.},
	year = {2002},
	langid = {english},
}

@article{benitez_investigating_2019,
	title = {Investigating the spin-orbit interaction in van der Waals heterostructures by means of the spin relaxation anisotropy},
	volume = {7},
	issn = {2166-532X},
	url = {https://doi.org/10.1063/1.5124894},
	doi = {10.1063/1.5124894},
	pages = {120701},
	number = {12},
	journal= {{APL} Materials},
	shortjournal = {{APL} Materials},
	author = {Benítez, L. Antonio and Sierra, Juan F. and Savero Torres, Williams and Timmermans, Matias and Costache, Marius V. and Valenzuela, Sergio O.},
	year = {2019},
}

@article{Milivojevic_giant_2024,
doi = {10.1088/2053-1583/ad59b4},
url = {https://doi.org/10.1088/2053-1583/ad59b4},
year = {2024},
month = {jun},
publisher = {IOP Publishing},
volume = {11},
number = {3},
pages = {035036},
author = {Milivojević, Marko and Gmitra, Martin and Kurpas, Marcin and Štich, Ivan and Fabian, Jaroslav},
title = {Giant asymmetric proximity-induced spin–orbit coupling in twisted graphene/{SnTe} heterostructure},
journal = {2D Materials}
}

@article{zutic_proximitized_2019,
	title = {Proximitized materials},
	volume = {22},
	issn = {1369-7021},
	url = {https://www.sciencedirect.com/science/article/pii/S1369702118301111},
	doi = {10.1016/j.mattod.2018.05.003},
	pages = {85--107},
	journal = {Materials Today},
	shortjournal = {Materials Today},
	author = {Žutić, Igor and Matos-Abiague, Alex and Scharf, Benedikt and Dery, Hanan and Belashchenko, Kirill},
	date = {2019-01-01},
    year = {2019},
}

@article{naimer_twist-angle_2021,
	title = {Twist-angle dependent proximity induced spin-orbit coupling in graphene/transition metal dichalcogenide heterostructures},
	volume = {104},
	url = {https://link.aps.org/doi/10.1103/PhysRevB.104.195156},
	doi = {10.1103/PhysRevB.104.195156},
	pages = {195156},
	number = {19},
	journal = {Physical Review B},
	shortjournal = {Phys. Rev. B},
	author = {Naimer, Thomas and Zollner, Klaus and Gmitra, Martin and Fabian, Jaroslav},
	date = {2021-11-30},
    year = {2021},
}

@article{gmitra_trivial_2016,
	title = {Trivial and inverted Dirac bands and the emergence of quantum spin Hall states in graphene on transition-metal dichalcogenides},
	volume = {93},
	url = {https://link.aps.org/doi/10.1103/PhysRevB.93.155104},
	doi = {10.1103/PhysRevB.93.155104},
	pages = {155104},
	number = {15},
	journal = {Physical Review B},
	shortjournal = {Phys. Rev. B},
	author = {Gmitra, Martin and Kochan, Denis and Högl, Petra and Fabian, Jaroslav},
	date = {2016-04-05},
    year = {2016},
}

@article{wang_strong_2015,
	title = {Strong interface-induced spin–orbit interaction in graphene on {WS}$_2$},
	volume = {6},
	rights = {2015 The Author(s)},
	issn = {2041-1723},
	url = {https://www.nature.com/articles/ncomms9339},
	doi = {10.1038/ncomms9339},
	pages = {8339},
	number = {1},
	journal = {Nature Communications},
	shortjournal = {Nat Commun},
	author = {Wang, Zhe and Ki, Dong-Keun and Chen, Hua and Berger, Helmuth and {MacDonald}, Allan H. and Morpurgo, Alberto F.},
	date = {2015-09-22},
	langid = {english},
    year = {2015},
}

@article{wang_origin_2016,
	title = {Origin and Magnitude of `Designer' Spin-Orbit Interaction in Graphene on Semiconducting Transition Metal Dichalcogenides},
	volume = {6},
	url = {https://link.aps.org/doi/10.1103/PhysRevX.6.041020},
	doi = {10.1103/PhysRevX.6.041020},
	pages = {041020},
	number = {4},
	journal = {Physical Review X},
	shortjournal = {Phys. Rev. X},
	author = {Wang, Zhe and Ki, Dong-Keun and Khoo, Jun Yong and Mauro, Diego and Berger, Helmuth and Levitov, Leonid S. and Morpurgo, Alberto F.},
	date = {2016-10-26},
    year = {2016},
}

@article{wakamura_strong_2018,
	title = {Strong Anisotropic Spin-Orbit Interaction Induced in Graphene by Monolayer {WS}$_2$},
	volume = {120},
	url = {https://link.aps.org/doi/10.1103/PhysRevLett.120.106802},
	doi = {10.1103/PhysRevLett.120.106802},
	pages = {106802},
	number = {10},
	journal = {Physical Review Letters},
	shortjournal = {Phys. Rev. Lett.},
	author = {Wakamura, T. and Reale, F. and Palczynski, P. and Guéron, S. and Mattevi, C. and Bouchiat, H.},
	date = {2018-03-09},
    year = {2018},
}

@article{oyedele_pdse2_2017,
	title = {{PdSe}$_2$: Pentagonal Two-Dimensional Layers with High Air Stability for Electronics},
	volume = {139},
	issn = {0002-7863, 1520-5126},
	url = {https://pubs.acs.org/doi/10.1021/jacs.7b04865},
	doi = {10.1021/jacs.7b04865},
	shorttitle = {{PdSe} $_{\textrm{2}}$},
	pages = {14090--14097},
	number = {40},
	journal = {Journal of the American Chemical Society},
	shortjournal = {J. Am. Chem. Soc.},
	author = {Oyedele, Akinola D. and Yang, Shize and Liang, Liangbo and Puretzky, Alexander A. and Wang, Kai and Zhang, Jingjie and Yu, Peng and Pudasaini, Pushpa R. and Ghosh, Avik W. and Liu, Zheng and Rouleau, Christopher M. and Sumpter, Bobby G. and Chisholm, Matthew F. and Zhou, Wu and Rack, Philip D. and Geohegan, David B. and Xiao, Kai},
	date = {2017-10-11},
	langid = {english},
    year= {2017}
}

@article{sun_electronic_2015,
	title = {Electronic, transport, and optical properties of bulk and mono-layer {PdSe}$_2$},
	volume = {107},
	issn = {0003-6951},
	url = {https://doi.org/10.1063/1.4933302},
	doi = {10.1063/1.4933302},
	pages = {153902},
	number = {15},
	journal = {Applied Physics Letters},
	shortjournal = {Applied Physics Letters},
	author = {Sun, Jifeng and Shi, Hongliang and Siegrist, Theo and Singh, David J.},
	date = {2015-10-13},
    year={2015}
}

@article{lu_layer-dependent_2020,
author = {Lu, Li-Syuan and Chen, Guan-Hao and Cheng, Hui-Yu and Chuu, Chih-Piao and Lu, Kuan-Cheng and Chen, Chia-Hao and Lu, Ming-Yen and Chuang, Tzu-Hung and Wei, Der-Hsin and Chueh, Wei-Chen and Jian, Wen-Bin and Li, Ming-Yang and Chang, Yu-Ming and Li, Lain-Jong and Chang, Wen-Hao},
title = {Layer-Dependent and In-Plane Anisotropic Properties of Low-Temperature Synthesized Few-Layer {PdSe}$_2$ Single Crystals},
journal = {ACS Nano},
volume = {14},
number = {4},
pages = {4963-4972},
year = {2020},
doi = {10.1021/acsnano.0c01139},
}

@article{yu_direct_2020,
	title = {Direct Observation of the Linear Dichroism Transition in Two-Dimensional Palladium Diselenide},
	volume = {20},
	issn = {1530-6984},
	url = {https://doi.org/10.1021/acs.nanolett.9b04598},
	doi = {10.1021/acs.nanolett.9b04598},
	pages = {1172--1182},
	number = {2},
	journal = {Nano Letters},
	shortjournal = {Nano Lett.},
	author = {Yu, Juan and Kuang, Xiaofei and Gao, Yuanji and Wang, Yunpeng and Chen, Keqiu and Ding, Zhongke and Liu, Jia and Cong, Chunxiao and He, Jun and Liu, Zongwen and Liu, Yanping},
	date = {2020-02-12},
    year={2020}
}

@article{xu_unravelling_2024,
	title = {Unravelling the origin of thermal anisotropy in {PdSe}$_2$},
	volume = {11},
	issn = {2053-1583},
	url = {https://doi.org/10.1088/2053-1583/ad64e3},
	doi = {10.1088/2053-1583/ad64e3},
	number = {4},
	urldate = {2025-10-28},
	journal = {2D Mater.},
	author = {Xu, Kai and Armesto, Luis Martínez and Světlík, Josef and Sierra, Juan F and Marinova, Vera and Dimitrov, Dimitre and Goñi, Alejandro R and Krysztofik, Adam and Graczykowski, Bartlomiej and Rurali, Riccardo and Valenzuela, Sergio O and Reparaz, Juan Sebastián},
	month = jul,
	year = {2024},
	pages = {045006},
}

@article{tombros_electronic_2007,
	title = {Electronic spin transport and spin precession in single graphene layers at room temperature},
	volume = {448},
	rights = {2007 Springer Nature Limited},
	issn = {1476-4687},
	url = {https://www.nature.com/articles/nature06037},
	doi = {10.1038/nature06037},
	pages = {571--574},
	number = {7153},
	journal = {Nature},
	author = {Tombros, Nikolaos and Jozsa, Csaba and Popinciuc, Mihaita and Jonkman, Harry T. and van Wees, Bart J.},
	date = {2007-08},
    year= {2007},
	langid = {english},
}

\end{document}